\documentclass[reqno]{article}
\RequirePackage{vmargin,latexsym,amssymb}
\setpapersize{A4}

\newcommand{\gC}{\Gamma}
\newcommand{\gd}{\delta}

\newcommand{\gep}{\epsilon}

\newcommand{\gT}{\Theta}

\newcommand{\gy}{\eta}

\newcommand{\di}{\partial}

\newcommand{\bgC}{\bar{\gC}}

\newcommand{\lb}[1]{\label{#1}}
\newcommand{\Eq}[1]{(\ref{#1})}

\renewcommand{\[}{\begin{eqnarray}}
\renewcommand{\]}{\end{eqnarray}}
\newcommand{\nn}{\nonumber}
\newcommand{\non}{\nonumber \\ }

\begin{document}
\arraycolsep3pt
\thispagestyle{empty}
\begin{flushright} hep-th/9907170\\
                   KCL-MTH-99-30
\end{flushright}
\vspace*{2cm}
\begin{center}
 {\LARGE \sc A Calibration Bound for\\[2ex]
             the M-Theory Fivebrane}\\
 
 \vspace*{1cm}
 {\sl
     Oliver B\"arwald\footnote[1]{Supported by the EC under TMR 
     contract ERBFMBICT972717,},
     Neil D.\ Lambert and
     Peter C.\ West\\
oliver, lambert, pwest@mth.kcl.ac.uk\\
 \vspace*{6mm}
     Department of Mathematics, King's College London\\
     Strand, London WC2R 2LS, Great Britain\\
 \vspace*{6mm}}
{July 21, 1999}\\
\vspace*{1cm}
\begin{minipage}{11cm}\footnotesize
  \textbf{Abstract:} We construct a covariant bound on the energy-momentum
of the M-fivebrane which is saturated by all  supersymmetric configurations.
This leads to a generalised notion of a calibrated geometry for M-fivebranes
when the worldvolume gauge field is non-zero. The generalisation relevant for 
D$p$-branes is also given.
\end{minipage}

\end{center}

\section{Introduction}

Supersymmetric soliton solutions have been at the centre of a great deal of
study. In general these solitons are stable due to their saturation of a 
topological bound which is a consequence of the existence of preserved
supersymmetries and an exact 
force cancellation between the various
fields.  In the case of a non-Abelian monopole for example, the Bogomol'nyi
condition relates the scalar Higgs field to the magnetic field. The 
quantum corrections to these states are also under greater control 
and therefore their properties can often 
be deduced in the complete quantum theory. 

In recent years the focus in String Theory has moved to branes where M-Theory
provides an elegant and unifying framework. M-Theory possess just two branes
in flat spacetime and the M-fivebrane is particularly
interesting since it has a gauge field on its worldvolume. In contrast to
other branes, this gauge field is a self-dual two-form. 
The various D$p$-branes and the NS-fivebrane in ten-dimensional  II string 
theory are related to the 
M-fivebrane by dimensional reduction and T-duality. 

The scalar fields that occur in the low energy dynamics describe the way a
brane is embedded in spacetime and as a consequence the corresponding 
Bogomol'nyi conditions will determine the geometry of this embedding.
In studies of brane solitons 
in the absence of a gauge fields it has been found that the resulting
geometries have an elegant interpretation  as
calibrated manifolds \cite{BeBeSt95,BBMOOY96,GibPap98,GaLaWe98a,AcOFSp98}. 

This has led to a search for a generalised notion of calibrations to include 
solitons with active worldvolume gauge fields \cite{GaLaWe98b}. 
In \cite{GuPa99,GuPaTo99}
the notion of a calibration was extended to cases where there is a background 
spacetime
gauge field. In this
paper we will use the covariant superembedding approach 
to construct a general Bogomol'nyi bound on the energy-momentum of 
an M-fivebrane. This bound  naturally suggests a generalised definition of a 
calibration for cases when the worldvolume gauge field is present. Finally
we mention how this bound can be dimensionally reduced to the worldvolume of
a D$p$-brane.

\section{An Energy-Momentum Bound}

An M-fivebrane is described by its embedding coordinates $X^{\underline n}$,
$\underline n = 0,1,2,...,10$, and $\Theta^{\underline \alpha}$, 
$\underline \alpha = 1,2,3,...,32$, as well as a two-form gauge field
$B_{mn}$, $m,n=0,1,2,...,5$, with field strength $H=dB$ 
which lives on its six-dimensional worldvolume.
In static gauge one identifies the worldvolume coordinates with the first
six coordinates of spacetime and sets half of the Fermions to zero. 
This leaves five scalar modes $X^{n'}$, $n'=6,7,8,9,10$, 
describing the transverse coordinates and sixteen Fermions 
$\Theta_\alpha^{\ i}$. These are $USp(4)$ symplectic Majorana-Weyl 
spinors with $\alpha=1,2,3,4$, $i=1,2,3,4$. In what follows we will
omit the spinor indices for simplicity.
  
In the superembedding formalism the residual supersymmetry  
transformation after setting half of the Fermions to zero takes the form 
\cite{GaLaWe98a,GaLaWe98b}
\[
\gd \gT = \gep(1-\gC)\lb{susy},
\]
where $\gC^2=1$.  We can therefore define a projector 
$P_\gT\equiv \frac12 (1-\gC)$ whose kernel consists of the preserved 
supersymmetries of a given bosonic field configuration.
Out of \Eq{susy} one can construct a manifestly positive definite object
\[
||\gep(1-\gC)||^2= \gep(1-\gC)(1-\gC)^\dagger \gep^\dagger \ge 0, \lb{prebound}
\]
which on physical grounds is expected to lead to a bound on the energy
involving the central charges of the fivebrane supersymmetry algebra.

Evaluating the bound of \Eq{prebound} one finds that the energy does not
naturally occur unless $H_{mnp}=0$. As such it is difficult to interpret
the bound arising from the variation of the fermion of equation \Eq{susy}.
A way out of this dilemma is to recognise that \Eq{prebound}
contains a hidden ambiguity; the $\Theta$ arise as
half of the Grassmann coordinates of the background superspace and since 
these are merely coordinates, we are free to make another choice.
The new Fermions will have a different supervariation and therefore lead to
a new bound. 
 
We start by performing  a field redefinition
\[
\Psi = \gT M\ ,
\]
for some invertible, field dependent matrix $M$. The new Fermions
$\Psi$ will transform as
\[
\gd \Psi = \gd \gT M + \cdots= \gep(1-\gC)M +\cdots \ ,
\]
where the ellipses denote terms which vanish after we set the Fermions to
zero. From \Eq{prebound}
we are led to the positive quantity
\[
||\gep(1-\gC)M||^2= \gep(1-\gC)MM^\dagger(1-\gC)^\dagger\gep^\dagger \ge 0
\lb{genbound}\ .
\]
Our aim is now to find a matrix $M$ such that \Eq{genbound} {\em does}
give a natural bound on the energy for a generic fivebrane configuration.
{}From this new bound we will be able to derive a notion of a generalised
calibration which holds in the presence of worldvolume gauge fields.

We now turn to the technical task of determining the matrix $M$ and
the exact form of the resulting bound. First let us describe in more
detail the M-fivebrane dynamics \cite{HoSeWe97a}. These were derived from the
superembeding formalism when applied to the M-fivebrane \cite{HowSez97}. 
The physical three-form 
field is $H$, which is closed. However, it is constrained to satisfy a
non-linear self-duality condition. To describe this condition it is
helpful to introduce a three-form $h_{abc}$ which is self-dual
\[
h_{abc}=\frac{1}{3!} \gep_{abcdef} h^{def},
\]
where we used tangent frame indices $a,b, \ldots$. These are related to the
world indices $m,n,\ldots$, by the usual vielbein $e_m^{\ a}$ 
constructed from the
induced metric which in static gauge is given by
\[
g_{mn} = \eta_{mn} + \partial_m X^{a'}\partial_n X^{b'} \gd_{a'b'}.
\]
Later we will need the explicit connection associated with this metric
which is given by
\[
\gC_{mn}{}^p = \di_m \di_n X^{a'} \di_q X^{b'} g^{qp} \gd_{a'b'}.
\]
Self-duality
implies that the only covariant tensors one can construct out of $h$
are given by
\[
k_{ab}   &\equiv& h_a{}^{cd}h_{bcd}\nn,\\
(kh)_{abc} &\equiv& k_a{}^d h_{dbc},\\
Q        &\equiv& 1- \frac23 k^{ab}k_{ab},\nn
\]
since $(k^2)_a^{\ b}=\frac16 ({\rm tr}k^2)\delta_a^{\ b}$.
One then defines the physical (closed) three-form field $H_{abc}$ as
\[
H_{abc} \equiv \frac{1}{Q}(h_{abc} + 2 (kh)_{abc})\ .
\]
The self-duality condition on $h_{abc}$ now implies that $(kh)_{abc}$ is
anti-self-dual resulting in 
\[
\star H^{abc} \equiv \frac{1}{3!}\gep^{abcdef} H_{def} =
\frac{1}{Q}(h^{abc} - 2 (kh)^{abc}).
\]
With these definitions the energy-momentum  can be written as \cite{us}
\[
T^{0a} =   \frac{2-Q}{Q} \gy^{0a} -
  \frac{4k^{0a}}{Q}\ ,
\lb{EMtensor}
\]
and in particular the energy is, for a static configuration, given by 
\[
E = {2-Q\over Q} + {4\over Q}k^{00}\ .
\lb{Energy}
\]
which agrees with the action formulation \cite{BLNPST97b}.

The worldvolume $\gC$-matrices are
the pull-backs of the flat eleven-dimensional $\bgC$-matrices 
and in static gauge are given as
\[
\gC_m =\partial_m X^{\underline n}\bar\Gamma_{\underline n}
= \gd_m{}^n \bgC_n + \partial_m X^{n'} \bgC_{n'}.
\]
But for our calculation it will be more convenient to use these
$\Gamma$-matrices referred to tangent frame $\Gamma_a = e_a^{\ m}\Gamma_m$;
here they obey the standard property $\{\gC_a,\gC_b\} = 2 \gy_{ab}$ 
and as a consequence also the duality relation
\[
\gC^{a_1 \ldots a_n} = \frac{1}{(6-n)!} (-1)^{\frac{n(n+1)}{2}}
\gep^{a_1\ldots a_n a_{n+1} \ldots a_6} \gC_{012345} \gC_{a_{n+1} \ldots a_6}.
\]
We may now write the matrix $\gC$ that appears in the supersymmetry 
projector $P_\Theta$ \Eq{susy}
as \cite{HoSeWe97a} 
\[
\gC = \gC_{(0)} + \gC_{(h)} \equiv - \frac{1}{6!} \gep^{abcdef}
\gC_{abcdef}+
\frac13 h^{abc} \gC_{abc}\ .
\] 
{}Furthermore, the M-fivebrane itself preserves supersymmetries such that
$\epsilon (1-\bar\Gamma_{012345}) = 0$. 

Returning to the matrix $M$, the most general covariant form that can be 
constructed from
the fields of the M-fivebrane is given by
\[
M \equiv a + b\gC_{(0)} + c \gC_{(h)} + d\gC_{(kh)},
\]
where $\gC_{(kh)} \equiv \frac13 (kh)^{abc} \gC_{abc}$ and $a,b,c$ and
$d$ are functions of $Q$ which are to be determined.  To proceed we
need the basic matrices in a more explicit way. 
Using self-duality of $h$,
anti-self-duality of $(kh)$ and the duality relation for the
$\gC$-matrices one finds
\[
\gC_{(0)} &=& - \gC_{012345},\nn\\
\gC_{(h)} &=& h^{0ij} \gC_{0ij} (1+ \gC_{012345}),\\
\gC_{(kh)} &=& (kh)^{0ij} \gC_{0ij} (1- \gC_{012345}),\nn
\]
here $i,j,\ldots$ denote tangent indices running from 1 to
5. We need three additional identities
\[
(h^{0ij}\gC_{0ij})^2 &=& 2 ( k^{00} + k_i{}^0 \gC_{012345} \gC^{0i}),\non
((kh)^{0ij}\gC_{0ij})^2 &=& \frac{1-Q}{2}(k^{00} - k_i{}^0 \gC_{012345} \gC^{0i}),\\
h^{0ij}\gC_{0ij}(kh)^{0kl}\gC_{0kl}&=&\frac12(Q-1).\nn
\]
A short calculation then reveals
\[
(1-\gC)M = (a-b)(1-h^{0ij} \gC_{0ij})(1+\gC_{012345})
+ d(2(kh)^{0ij}\gC_{0ij} + 1-Q)(1-\gC_{012345}) \lb{poldM}.
\]
We are only concerned with static configurations which implies
\[
(\gC_{012345})^\dagger = \gC_{012345} \quad \mbox{ and } \quad
(\gC_{0ij})^\dagger = \gC_{0ij}.
\]
{}Furthermore one has $\{\gC_{0ij},\gC_{012345}\}=0$. Hence we find the
conjugate of \Eq{poldM} to be
\[
M^\dagger(1-\gC^\dagger) = (a-b)(1+\gC_{012345})(1-h^{0ij} h_{0ij})
+ d (1-\gC_{012345})(2 (kh)^{0ij}\gC_{0ij} + 1-Q).
\]
Evaluating the product and sorting the resulting terms gives
\[
(1-\gC)M M^\dagger (1-\gC^\dagger) &=&
\hphantom{+}2((a-b)^2 + d^2(1-Q)^2) \non
&&+2((a-b)^2 + d^2(1-Q)) 2 (k^{00}- k_i{}^0 \gC^{0i})\non
&&+2(-(a-b)^2 + d^2(1-Q)) 2 (k^{00}- k_i{}^0 \gC^{0i}) \gC_{012345}\\
&&+2((a-b)^2 - d^2(1-Q)^2)\gC_{012345}\non
&&+2(a-b)^2 (-2 h^{0ij}\gC_{0ij})+2d^2(1-Q) 4 (kh)^{0ij}\gC_{0ij}\nn.
\]
Making the choice $d^2=\frac{1}{1-Q}(a-b)^2$ leads to
\[
(1-\gC)M M^\dagger (1-\gC^\dagger) = 2(a-b)^2 Q
(E-\gC^{0i} T_i{}^0 - \gC_{0} -2 \star\! H^{0ij} \gC_{0ij}),
\]
where we have identified the components of the energy-momentum tensor 
\Eq{EMtensor}. 
So for this choice of constants we obtain the following bound on the
energy
\[
\gep(E-\gC^{0i} T_i{}^0) \gep^\dagger \ge \gep(\gC_{(0)}
 +2 \star\!H^{0ij} \gC_{0ij})\gep^\dagger \ ,
\label{EMb}
\] 
which is the same as found in \cite{GaLaWe98b} if we identify $t_i=P_i$.
It is gratifying to see that the three-form automatically appears in the
guise of the dual of the physical field $H_{abc}$. Note that the  left
hand side of \Eq{EMb} is simply 
\[
\bar \epsilon \Gamma^m P_m\epsilon^\dag\ ,
\]
where $\bar\epsilon = \epsilon\Gamma_0$ and $P_m = T_{m}^{\ 0}$ is the
six-momentum density. Spatial integration of equation \Eq{EMb} leads to a
manifestly covariant form for the bound in terms of
the Noether charges.  

In an earlier work \cite{us} we found that the associated energy
momentum tensor is {\em covariantly} conserved
\[
\nabla_m T^{mn} =0.
\]
We are now going to show that this tensor gives rise to a tensor density
\[
\tilde{T}^{mn} = \sqrt{-g}T^{mn},
\]
which is conserved in the {\em flat space} sense
\[
\di_m \tilde{T}^{mn} =0.
\]
Thus the total energy and momentum of the M-fivebrane are indeed conserved 
quantities.

The equation of motion for the scalar fields of the fivebrane in
static gauge and for a flat background is given by \cite{HoSeWe97a}
\[
G^{mn} \nabla_m \nabla_m X^{a'} =0.\lb{EOM}
\]
Here $G^{mn}=QT^{mn}$ \cite{us} so this can be rewritten as
\[
T^{mn} \nabla_m \nabla_n X^{a'} = \nabla_m (T^{mn} \nabla_n X^{a'})=0,
\]
where we used the fact that $T^{mn}$ is covariantly conserved. 
Using standard tensor analysis 
we find that the equation of motion can be written as
\[
\frac{1}{\sqrt{-g}} \di_m (\sqrt{-g} T^{mn} \di_n X^{a'})=
\frac{1}{\sqrt{-g}} \di_m (\sqrt{-g} T^{mn})\di_n X^{a'}
+ T^{mn} \di_m \di_n X^{a'}  =0.
\]
Hence to prove our initial statement we have to show that the second
term vanishes. This is equivalent to showing that the equation of
motion implies $G^{mn} \di_m \di_n X^{a'}=0$. 
Starting from \Eq{EOM} we calculate
\[
\begin{array}{rcl}
G^{mn}\nabla_m \di_n X^{a'} &=& G^{mn}\di_m \di_n X^{a'}
- G^{mn} \gC_{mn}{}^p \di_p X^{a'},\\
&=&G^{mn}\di_m \di_n X^{a'}
- G^{mn} \di_m \di_n X^{b'} \di_q X^{c'} g^{qp} \gd_{b'c'} \di_p X^{a'},\\
&=& G^{mn} \di_m \di_n X^{b'} ( \gd_{b'}{}^{a'} - \di_q X^{c'} \di_p
X^{a'}
g^{qp} \gd_{b'c'} ),\\
&=& G^{mn} \di_m \di_n X^{b'} M_{b'}{}^{a'}
\end{array}
\]
A short calculation reveals that the inverse of the matrix
$M_{b'}{}^{a'}$ exists and is given by
\[
(M^{-1}){}_{a'}{}^{b'} = \gd_{a'}{}^{b'} + \di_m X^{c'} \di_n X^{b'}
\gd_{c'a'} \gy^{mn} .
\]
Thus it follows that $G^{mn}\nabla_m \di_n X^{a'} =0$ if and only if 
$G^{mn} \di_m \di_n X^{a'}=0$. 

Returning to the bound it turns out that for any choice of $a,b,c$ and
$d$, the matrix $(1-\gC)M$ squares to a multiple of itself
\[
((1-\gC)M)^2=2(a-b+d(1-Q))(1-\gC)M.
\]
Hence the variation of the new Fermions under worldvolume supersymmetry
can be rewritten in the form
\[
\delta\Psi = \epsilon P_\Psi \ ,
\]
where we defined 
\[
P_{\Psi} \equiv \frac{1}{2(a-b+d(1-Q))} (1-\gC)M.
\]
Choosing $d=\frac{1}{\sqrt{1-Q}}(a-b)$ as before we find from \Eq{poldM}
\[
P_\Psi &=& \frac{1}{2(1+\sqrt{1-Q})}\bigg((1-h^{0ij} h_{0ij})(1+\gC_{012345})
\nn\\
&&+ \Big(\frac{2}{\sqrt{1-Q}} (kh)^{0ij}\gC_{0ij}+ \sqrt{1-Q}\Big)(1-\gC_{012345})\bigg),
\]
the remaining constants drop out and we are led to a unique new
projector. If we express $P_\Psi$ in a more familiar
way we find
\[
2 P_\Psi = 1- \frac{1- \sqrt{1-Q}}{1+ \sqrt{1-Q}}\left(\gC_{(0)} +
  \frac{1}{1+\sqrt{1-Q}}\gC_{(h)}
+ \frac{2}{\sqrt{1-Q}(1+\sqrt{1-Q})} \gC_{(kh)}\right).
\]
In principle one can now forget about $M$ and simply take $P_\Psi$ to
be the new supersymmetry projector associated with the new set of
{}Fermions $\Psi$. A short calculation gives two relations fulfilled by
$P_\Psi$ and $P_\gT$
\[
P_\gT P_\Psi = P_\Psi,\quad P_\Psi P_\gT= P_\gT.
\]
These relations imply that $P_\gT$ and $P_\Psi$ have the same zero
modes, recall that they act from the right, and that the image
of one is related by an invertible, linear transformation to the image of
the other. Therefore the Bogomol'nyi conditions for the preservation of 
supersymmetry are identical. Although a similar equation was 
obtained in \cite{BLNPST97b} in
terms of $P_\Theta$ and the supersymmetry projector $P_{PST}$ of the
action formulation, it is easily seen that $P_{PST}\ne P_\Psi$
since the former is Hermitian whereas the latter is not.

Our final task is to ensure that 
$M$ is invertible. In fact it is easy to establish that 
this is generically the case for any values of
$a,b,c,d$. A particularly simple  choice is to take 
$a=0,b=1,c=0$ so that  $d=(1-Q)^{-\frac12}$ and
\[
M = \gC_{(0)} 
+ {1\over \sqrt{1-Q}} \gC_{(kh)}\ ,
\]
with $M^2=1$.

\section{Generalised Calibrations}

In the previous section we showed that there is a bound on the energy-momentum
of a static M-fivebrane configuration which is saturated precisely when
supersymmetry is preserved. For the purely scalar case it is know that the
right hand side of the bound corresponds to a calibrating form and that
supersymmetric states can be interpreted as an M-fivebrane wrapped on a
calibrated submanifold of eleven-dimensional spacetime. In this section we
would like to find a similar spacetime interpretation in terms of a 
``generalised calibration'' when $H \ne 0$. 

Let us restrict our attention to static
configurations that have a rest frame where  $P^i=T^{0i}=0$. We are also
only interested in the spatial section of the M-fivebrane worldvolume. 
Let us introduce the five-form $\varphi$ and two-form $\chi$ defined as
\[
\varphi &=& 
\frac{1}{5!}\epsilon\bar\Gamma_{0\mu_1...\mu_5}\epsilon^\dag
dx^{\mu_1}\wedge ...\wedge dx^{\mu_5}\nn\ ,\\
\chi &=& -\frac12\epsilon\bar \Gamma_{0\mu_1\mu_2}
\epsilon^\dag dx^{\mu_1}\wedge dx^{\mu_2}\nn\ ,
\]
where $\mu_1,\mu_2,\mu_3...=1,2,3,...,10$
and we choose to normalise our spinors such that 
$\epsilon\epsilon^\dag=1$.
These forms are defined over the entire 
ten-dimensional space in which the M-fivebrane sits. 
Furthermore, since they are constant, they are
closed. The bound on the total energy now can be written as
\[
{\cal E} \ge \int_{M} \ {}^\star \varphi + H\wedge {}^\star\chi \ ,
\label{Ebound}
\]
where ${}^\star$ denotes the pull-back to the spatial part of the M-fivebrane 
worldvolume $M$. Here we see that the role of $\chi$ is to pick out
preferred three-cycles in the worldvolume over which the 
flux of $H$ is measured. Note that only the spatial components
of $H$ contribute to ${\cal E}$.
 
This very suggestive form for the bound motivates the following
definition of a generalised calibration, valid for an arbitrary background
spacetime: A generalised calibration consists
of a closed five-form $\varphi$,  together with a closed two-form $\chi$
such that, for all submanifolds $M$ 
with a three-form $H$ defined on it, 
\[
E\  {\rm d}vol \ge {}^\star \varphi + H\wedge {}^\star\chi\ ,
\lb{calform}
\]
when evaluated on any tangent plane. Here
${\rm d}vol$ is the volume form and $E$ is given by \Eq{Energy}. 
A particular pair $(M,H)$ consisting of a
five-dimensional submanifold $M$ and a three-form $H$ is then said to be 
calibrated by $(\varphi,\chi)$
if the inequality is saturated on all tangent planes of $M$.
Clearly for $H=0$ then $E=1$ and \Eq{calform} reduces to the standard
definition of a calibrating form.  
This leads to the theorem: A calibrated
pair $(M,H)$ has minimal energy $\cal E$ with respect to 
all other pairs $(M',H')$
such that $M-M'=\partial {N}$, for some manifold $N$
and $H=H'$ on $\partial M=\partial M'$. This is
easily established from the closure of $\varphi, \chi$  and 
Stokes theorem 
as in the analogous proof for standard calibrations
\[
{\cal E} &=& \int_M  {}^\star \varphi + H\wedge {}^\star\chi \nn\ ,\\
&=&  \int_{M} {}^\star \varphi 
+ \int_{\partial M}B\wedge {}^\star\chi\nn\ ,\\ 
&=&\int_{M'} {}^\star \varphi +\int_{\partial {N}} {}^\star\varphi
+ \int_{\partial M}(B'-d\Lambda)\wedge {}^\star\chi\nn\ ,\\
&=& \int_{M'}{}^\star \varphi +\int_{N}{}^\star d\varphi
+ \int_{\partial M'}B'\wedge {}^\star\chi\nn\ ,\\ 
&=&\int_{M'} {}^\star \varphi 
+ H'\wedge {}^\star\chi\ \  \le {\cal E}'\nn\ ,\\ 
\]
where we have written $B'=B+d\Lambda$ on $\partial M$.
Thus a
calibrated pair minimises the energy and therefore 
solves the M-fivebrane equations of motion. 

Although the right hand side
of \Eq{Ebound} seems quite simple the form for $E$ given in 
equation \Eq{Energy} is rather complicated. However the this form is
necessary for the existence of calibrated M-fivebranes and we therefore
suspect that there is  a deeper geometrical significance to $E$ which would be
interesting to understand. Furthermore  the presence
of preserved supersymmetries  implies the existence of reduced holonomy
in spacetime. However if the preserved supersymmetries and in particular
their Bogomol'nyi conditions rely on gauge fields, either on the worldvolume
or in spacetime, then the reduced holonomy is not with respect to the
Levi-Civita connection but rather a connection that sees the relevant
gauge structure. It might be interesting to find a classification of 
all reduced holonomies and the geometries that result from the corresponding
generalised calibrations.

We may further generalise the notion of a calibration to include the
case where $d\varphi={\cal F}\ne0$ as in \cite{GuPa99,GuPaTo99} 
and also for $dH\ne 0$.
These cases correspond to a non-zero four-form field strength $G=dC$ in
eleven dimensions. In this case we may write $H=dB+{}^\star C$ where 
${}^\star C$ is the pull-back of $C$ and $dH={}^\star G$ \cite{HowSez97}.
The above argument is altered to
\[
{\cal E}\ 
\le\  {\cal E}' + \int_{N} {}^\star({\cal F}+G\wedge\chi)\ .
\lb{wzbound}
\]
Following \cite{GuPa99,GuPaTo99} we may interpret the additional term as the
contribution of a Wess-Zumino term to the total energy. 
To see this we observe that the modified bound \Eq{wzbound} can be writen as
\[
{\cal E} + WZ \le \ {\cal E}' + WZ' \ ,
\]
where 
\[
WZ = -\int_{M} {}^\star(\Phi + C\wedge\chi)\ ,\quad
WZ' = -\int_{M'}{}^\star (\Phi + C\wedge\chi)\ ,
\]
and $\Phi$ is an static gauge potential  
defined by $d\Phi = {\cal F}$. 

In practice many soliton solutions of the M-fivebrane are $r$-branes 
where the scalars and 
three-form only depend on $q = 5-r$ of the spatial dimensions. In this case
the forms $\varphi$ and $\chi$ are more naturally interpreted as 
a $q$-form and $(q-3)$-form respectively
\[
\varphi&=& 
\epsilon\bar\Gamma_{01...r\mu_1...\mu_q}\epsilon^\dag
dx^{\mu_1}\wedge ... \wedge dx^{\mu_q}\ ,\nn\\
\chi &=& -\epsilon\bar\Gamma_{01...r\mu_1...\mu_{q-3}}\epsilon^\dag
dx^{\mu_1}\wedge...\wedge dx^{\mu_{q-3}}\ .\nn\\
\]
Furthermore we are now interested in the energy per unit $r$-volume and so only
the components of $H$ tangent to the $x^{r+1},...,x^5$ dimensions appear
in the bound.
The definition of a calibration follows analogously and it
is clear that the above theorem still holds for $q$-dimensional submanifolds
$M$.

There are already many examples of solitons on the M-fivebrane which are
supersymmetric with a non-zero $H$ \cite{HoLaWe98,GaLaWe98b,LamWes98c,
GaKhMaTo99,Gauntlett99,GoRaSiTo99}. 
These therefore provide examples of
generalised calibrations as defined above. Perhaps the simplest example
is the self-dual string soliton \cite{HoLaWe98} with $q=4$. This soliton has
\[
H_{01i} = {1\over4} \partial_i X^6\ ,\quad 
H_{ijk} = {1\over4}\epsilon_{ijkl}\partial_lX^6\ ,\quad
X^6 = a + {Q\over r^2}\ ,
\]
where $r^2 = x^ix^i$, $ i=2,3,4,5$
and all other fields vanish. 
The preserved supersymmetries satisfy $\epsilon \bar\Gamma_{016} 
=-\epsilon$ and
\[
\varphi = 
\epsilon\bar\Gamma_{01\mu\nu\lambda\rho}\epsilon^\dag
dx^{\mu}\wedge dx^{\nu}\wedge dx^{\lambda}\wedge dx^{\rho}\ ,\quad
\chi
= -\epsilon\bar\Gamma_{01\mu}\epsilon^\dag dx^\mu\ .
\]
The energy for this configuration is \cite{GaGoTo98} 
\[
E\ {\rm d}vol= (1 + \partial_iX^6\partial_i X^6)
dx^2\wedge dx^3\wedge dx^4\wedge dx^5\ ,
\]
whereas the  forms ${}^\star\varphi$ and $H\wedge {}^\star\chi$ are 
\[
{}^\star\varphi &=& \left(
1-\partial_iX^6\epsilon\bar\Gamma_{6i}\epsilon^\dag
\right)dx^2\wedge dx^3\wedge dx^4\wedge dx^5\ ,\nn\\
H\wedge {}^\star\chi &=&\left(
\partial_i X^6\partial_i X^6\epsilon\epsilon^\dag 
+ \partial_i X^6\epsilon\bar\Gamma_{6i}\epsilon^\dag 
\right)dx^2\wedge dx^3\wedge dx^4\wedge dx^5 \nn .\\
\] 
{}From this and the normalisation $\epsilon\epsilon^\dag =1$
we see that the calibration condition \Eq{calform} is indeed saturated.

Finally let us describe how this definition of a generalised calibration
is modified when we dimensionally reduce the M-fivebrane to D-fourbrane. 
To this end we consider configurations for which the $x^5$ direction is
an isometry. It was shown in \cite{HoSeWe97a} that the M-fivebrane
equations of motion reduce to those of a D-fourbrane 
(with a Dirac-Born-Infeld action)  if the direction $x^5$ is compactified.
Thus if we consider solitons which do not depend on $x^5$ and write for
$m,n=0,1,2,3,4$  \cite{HoSeWe97a}
\[
F_{mn} = {1\over4}H_{mn5}\ ,
\]
then we find the generalised calibration may be expressed as
\[
E\ {\rm d}vol \ge \varphi + F\wedge\chi + \star \tilde F\wedge \omega\ , 
\lb{BIbound}
\]
which now includes contributions from both electric and magnetic
charges.  Here $\star$ is the Hodge dual on the D-fourbrane,
$\omega=-\epsilon\bar\Gamma_{0\mu 5}\epsilon^\dag dx^\mu$ and
\cite{HoSeWe97b}
\[
\tilde F = {(1-\frac12{\rm tr} F^2) F+ F^3\over \sqrt{\det(1+F)}}\ .
\] 
Here $F$ appears with one raised and one lowered index and matrix
multiplication is understood. In terms of the D-fourbrane fields we
may write $E = -\sqrt{{\rm det}(1+F)}(g+F)^{00}$, where $(g+F)^{mn}$
is the matrix inverse of $g_{mn}+F_{mn}$. It follows from the
Dirac-Born-Infeld equation of motion that $d\star\tilde F=0$, as we
expect.  The generalisation to all D$p$-branes with a
Dirac-Born-Infeld effective action is clear. Equation \Eq{BIbound}
stills applies but now the generalised calibration depends on a
$p$-form $\varphi$, $(p-2)$-form $\chi$ and a one-form $\omega$. In
all these cases an analogous theorem exists for a bound on the energy.

\end{document}